\documentclass[aps,pra,twocolumn,groupedaddress,noshowpacs]{revtex4}
\usepackage{bm}
\usepackage{epsf}
\usepackage{amssymb}
\usepackage{amsmath}
\usepackage{graphicx}
\usepackage{rotating}
\usepackage{epsfig}
\usepackage{psfrag}
\usepackage{amsmath}
\usepackage{hyperref}
\usepackage{physics}
\usepackage{color}
\hypersetup{
    bookmarks=true,         
    unicode=false,          
    pdftoolbar=true,        
    pdfmenubar=true,        
    pdffitwindow=true,      
    pdftitle={My title},    
    pdfauthor={Author},     
    pdfsubject={Subject},   
    pdfcreator={Creator},   
    pdfproducer={Producer}, 
    pdfkeywords={keywords}, 
    pdfnewwindow=true,      
    colorlinks=false,       
    linkcolor=red,          
    citecolor=green,        
    filecolor=magenta,      
    urlcolor=cyan           
}

\DeclareMathAlphabet{\bi}{OML}{cmm}{b}{it}

\begin{document}
\def\G{{\cal G}}
\def\F{{\cal F}}
\def\ea{\textit{et al.}}
\def\bM{{\bm M}}
\def\bN{{\bm N}}
\def\bV{{\bm V}}
\def\bj{\bm{j}}
\def\bSig{{\bm \Sigma}}
\def\bLam{{\bm \Lambda}}
\def\bfeta{{\bf \eta}}
\def\bn{{\bf n}}
\def\d{{\bf d}}
\def \xy{$x$--$y$ }
\def\bP{{\bf P}}
\def\bK{{\bf K}}
\def\bk{{\bf k}}
\def\bkn{{\bf k}_{0}}
\def\bx{{\bf x}}
\def\bz{{\bf z}}
\def\bR{{\bf R}}
\def\br{{\bf r}}
\def\bu{{\bm u}}
\def\bq{{\bf q}}
\def\bp{{\bf p}}
\def\by{{\bf y}}
\def\bQ{{\bf Q}}
\def\bs{{\bf s}}
\def\bA{{\mathbf A}}
\def\bv{{\bf v}}
\def\b0{{\bf 0}}
\def\la{\langle}
\def\ra{\rangle}
\def\Im{\mathrm {Im}\;}
\def\Re{\mathrm {Re}\;}
\def\beq{\begin{equation}}
\def\eeq{\end{equation}}
\def\bea{\begin{eqnarray}}
\def\eea{\end{eqnarray}}
\def\bdm{\begin{displaymath}}
\def\edm{\end{displaymath}}
\def\bnab{{\bm \nabla}}
\def\Tr{{\mathrm{Tr}}}
\def\bJ{{\bf J}}
\def\bU{{\bf U}}
\def\bPsi{{\bm \delta\Delta}}
\def\mA {\mathrm{A}}
\def \R{R_{\mathrm{s}}}
\def \rhos{n_{\mathrm{s}}}
\def \rhon{\tilde{n}}
\def \Rd{R_{\mathrm{d}}}
\def \xy{three dimensional $XY\;$}
\def\sfrac{\textstyle\frac}
\def \leq{\mathrm{l.e.}}
\def\p{\partial}
\def\nn{\nonumber}
\def\m{\mu}
\def\n{\nu}
\title{On operator product expansion in the spin-orbit coupled bosonic system}
\author{Rajesh Kumar Gupta}\email{rajesh.gupta@iitrpr.ac.in}
\author{Siddhant Tiwari}\email{siddhant.22phz0010@iitrpr.ac.in}
\affiliation{Department of Physics, Indian Institute of Technology Ropar}

\begin{abstract}
Ultra-cold bosonic systems can be tuned to exhibit quantum phase transitions. For example, the Rabi-coupled bosonic system exhibits ferromagnetic and paramagnetic phases, whereas the spin-orbit-coupled system exhibits exciting phases such as supersolidity. The physics of these phases and phase transitions is very rich. It is an important topic of research to probe these phases and phase transitions using various tools in many-body physics. The operator product expansion (OPE) provides one such tool. It expresses the product of two separated operators as a series expansion of local operators. In this article, we will derive the OPE of two operators $\psi^\dagger_\sigma(\vec r)$ and $\psi_{\sigma'}(\vec r')$. More specifically, we look for the contact density term, which controls many of the universal physics of the underlying bosonic system.
\end{abstract}

\maketitle
\section{Introduction}
Ultra-cold atomic systems consisting of a mixture of Bose gases of different species provide an avenue to explore rich and exotic quantum phenomena. An example of such a system consists of a mixture of Bose atoms in doublet hyperfine states, also called pseudo-spinor bosons, with an interaction between the hyperfine states induced by a counter propagating laser beams along the $x$-axis. The pseudo-spinor system with spin-orbit coupling can be realized experimentally and has been the topic of extensive research in the last decade~\cite{lin2009synthetic, lin2011synthetic, lin2011spin, wu2016realization, bersano2018experimental}. An interesting feature of the system is that it exhibits distinct phases such as zero momentum, plane wave and supersolid phase~\cite{PhysRevLett.107.150403, li2012quantum}. The physics of these phases and the phase transition is a topic of great interest. 

The action of a spin-orbit coupled bosonic system is
\bea\label{Action.1}
S&=&\int dt\int d^3x\,\Big[\psi^\dagger_1\Big(i\p_t+\frac{\nabla_\perp^2}{2}+\frac{1}{2}(\p_x-i\kappa_0)^2\Big)\psi_1+\nn\\
&&\psi^\dagger_2\Big(i\p_t+\frac{\nabla_\perp^2}{2}+\frac{1}{2}(\p_x+i\kappa_0)^2\Big)\psi_2
-\frac{\Omega}{2}(\psi^\dagger_1\psi_2\nn\\
&&+\psi^\dagger_2\psi_1)-\frac{g}{2}(|\psi_1|^4+|\psi_2|^4)-\lambda|\psi_1|^2|\psi_2|^2\Big]\,.
\eea
Here $\Psi=\begin{pmatrix}\psi_{1}&\psi_2\end{pmatrix}^T$ diagonalizes the Pauli matrix $\sigma_z$. Also, $\kappa_0$ is the momentum transfer due to the laser, and $\Omega$ denotes the coupling of atoms with the laser. For $\kappa_0=0$, the system is called the Rabi or coherently coupled bosonic system and for $\kappa_0\neq 0$ it is called the Raman coupling (or spin-orbit coupled) situation. Typically, in such a model, we require that the intra-species coupling constant is positive, $g>0$, to make the Hamiltonian bounded from below, but no condition is imposed on the sign of the inter-species coupling constant, $\lambda$.

The spin-orbit coupled system exhibits translational invariance; however, the Galilean invariance along the $x$-direction is broken due to the presence of the momentum transfer $\kappa_0$. In the Rabi case, i.e. $\kappa_0=0$, the full Galilean invariance of the system is restored. This has an important consequence for any computation. The physical result of a computation depends on the choice of the frame in the spin-orbit coupled case, for example the S-matrix, which respects the symmetries of the underlying Hamiltonian. However, one expects the physical results for the Rabi case to be frame-independent. 

The single particle Hamiltonian is not diagonal in the field excitations labelled by $\psi_i$.
A useful basis to consider is the one where the single particle Hamiltonian is diagonal. The diagonalization gives the physical excitation, which will be called a dressed particle. Suppose we denote these excitation by the field operator $\alpha_\pm(\vec r,t)$. Then, consider the unitary transformation to define a new creation and annihilation operator as
\beq\label{UnitaryTrans.1}
\begin{pmatrix}\alpha_+(\vec k)\\\alpha_-(\vec k)\end{pmatrix}=\begin{pmatrix}u(\vec k)&v(\vec k)\\-v^*(\vec k)&u^*(\vec k)\end{pmatrix}\begin{pmatrix}\psi_1(\vec k)\\\psi_2(\vec k)\end{pmatrix}\,.
\eeq
where the matrix coefficients satisfy
\beq
|u(\vec k)|^2+|v(\vec k)|^2=1\,.
\eeq
Choosing $u(\vec k)=\cos\frac{\theta_k}{2}$ and $v(\vec k)=\sin\frac{\theta_k}{2}$, we can write the single particle Hamiltonian as
\begin{align}
&H_{sp}=\sum_{\vec k}\Big[\Big(\frac{\vec k^2}{2}+\frac{1}{2}\kappa_0^2+\frac{\Omega}{2}\sin\theta_k+k_x\kappa_0\cos\theta_k\Big)\alpha_+^\dagger(\vec k)\alpha_+(\vec k)\nn\\
&+\Big(\frac{\vec k^2}{2}+\frac{1}{2}\kappa_0^2-\frac{\Omega}{2}\sin\theta_k-k_x\kappa_0\cos\theta_k\Big)\alpha_-^\dagger(\vec k)\alpha_-(\vec k)\nn\\
&+(-k_x\kappa_0\sin\theta_k+\frac{\Omega}{2}\cos\theta_k)(\alpha_+^\dagger(\vec k)\alpha_-(\vec k)+\alpha_-^\dagger(\vec k)\alpha_+(\vec k))\Big]
\end{align}
Requiring the off-diagonal term to vanish, we obtain
\beq
\tan\theta_k=\frac{\Omega}{2k_x\kappa_0}\,.
\eeq
It is interesting to note that the angle $\theta_k$ explicitly depends on the sign of $k_x$.
We will focus on the region for $\theta$ lying between $0<\theta<\pi$.
From the above we can obtain the expression for $u$ and $v$ which are 
\bea
u(\vec k)=\sqrt{\frac{1}{2}+\frac{k_x\kappa_0}{\tilde\Omega}},\quad v(\vec k)=\sqrt{\frac{1}{2}-\frac{k_x\kappa_0}{\tilde\Omega}}\,,
\eea
where $\tilde\Omega=\sqrt{\Omega^2+4k_x^2\kappa_0^2}$. It is important to emphasize that the coefficients of the unitary transformation, $u(\vec k)$ and $v(\vec k)$, depend on the $x$-component of the momentum. This will have an important implication on the form of the scattering amplitude for $2\rightarrow 2'$ process, as we will see later.
Also, we note that for the coherent case, $\theta_k=\frac{\pi}{2}$ and the diagonal basis simplifies to
\beq
\alpha_+(\vec k)=\frac{1}{\sqrt{2}}(\psi_1(\vec k)+\psi_2(\vec k)),\, \alpha_-(\vec k)=\frac{1}{\sqrt{2}}(\psi_2(\vec k)-\psi_1(\vec k))\,.
\eeq
Thus, the single particle Hamiltonian becomes
\begin{align}
H_{sp}=&\sum_{\vec k}\Big[\Big(\frac{\vec k^2}{2}+\frac{1}{2}\kappa_0^2+\sqrt{\frac{\Omega^2}{4}+k_x^2\kappa_0^2}\Big)\alpha_+^\dagger(\vec k)\alpha_+(\vec k)\nn\\
&+\Big(\frac{\vec k^2}{2}+\frac{1}{2}\kappa_0^2-\sqrt{\frac{\Omega^2}{4}+k_x^2\kappa_0^2}\Big)\alpha_-^\dagger(\vec k)\alpha_-(\vec k)\Big]\,.
\end{align}
Note that the vacuum of the physical excitations generated by $\alpha_\pm^\dagger$ is also the vacuum of $\psi_{1/2}$.

A fundamental quantity of our interest in this article is the operator product expansion (OPE). OPE is a useful quantity to compute in a quantum field theory, which encodes physics at short distances or large $\vec {k} $ behaviour. The OPE analysis is also relevant in the context of the ultra-cold atomic system since it has provided insights into universal relations, such as the Tan's relation in both fermionic and bosonic systems~\cite{tan2005large,tan2008energetics, braaten2008exact, braaten2008universal, braaten2010short, Braaten1101}. In OPE, the product of two local operators separated by a short distance is expressed as a linear combination of local operators~\cite{PhysRev.179.1499}. It is an operator statement that holds for any correlation function or expectation value. Thus, given the two local operators $\psi^\dagger_\sigma(\vec R-\frac{1}{2}\vec r)$ and $\psi_{\sigma'}(\vec R+\frac{1}{2}\vec r)$, say inside a correlation function, the OPE expresses the product of the two as an infinite sum:
\beq
\psi^\dagger_\sigma(\vec R-\frac{1}{2}\vec r)\psi_{\sigma'}(\vec R+\frac{1}{2}\vec r)\stackrel{r\rightarrow 0}{=}\sum_{\mathcal O}C_{\sigma\sigma',n}(\vec r)\mathcal O_{\sigma\sigma',n}(\vec R)
\eeq
Here $O_{\sigma\sigma',n}(\vec R)$ is a local operator constructed out of creation and annihilation operators and $C_{\sigma\sigma',n}(\vec r)$ is called the Wilson coefficient. Wilson coefficients contain the information about the UV or short-distance physics.
For example, the asymptotic fall-off of the momentum distribution of atoms for the large wave number $\vec k$ in a many-body state $\ket{X}$ is given by
\begin{align}
&n_{\sigma\sigma'}(\vec k)=\int d^3R\int d^3r\,e^{i\vec k\cdot\vec r}\bra{X}\psi^\dagger_\sigma(\vec R-\frac{1}{2}\vec r)\psi_{\sigma'}(\vec R+\frac{1}{2}\vec r)\ket{X}\nn\\
&\stackrel{k\rightarrow\infty}{=}\sum_{\mathcal O}\Big(\int d^3r\,e^{i\vec k\cdot\vec r}C_{\sigma\sigma',n}(\vec r)\Big)\int d^3R\,\bra{X}\mathcal O_{\sigma\sigma',n}(\vec R)\ket{X}
\end{align}
Thus, the large $k$-dependence of the momentum distribution of density is determined by the small $\vec {r} $ OPE expansion and, in particular, the knowledge of the Wilson coefficient. Based on the above consideration, we aim to derive the OPE between two local operators in the spin-orbit coupled bosonic system. For the simplest operators $\psi^\dagger_\sigma(\vec R-\frac{1}{2}\vec r)$ and $\psi_{\sigma'}(\vec R+\frac{1}{2}\vec r)$, we expect to arrive at
\begin{align}
&\psi^\dagger_\sigma(\vec R-\frac{1}{2}\vec r)\psi_{\sigma'}(\vec R+\frac{1}{2}\vec r)\stackrel{r\rightarrow 0}{=}\psi^\dagger_\sigma\psi_{\sigma'}(\vec R)+\vec r\cdot\psi^\dagger_\sigma\overset{\leftrightarrow}{\nabla}\psi_{\sigma'}(\vec R)\nn\\
&-\frac{r}{8\pi}C^{\sigma\sigma'}_{\alpha\beta\gamma\delta}\psi_\alpha^\dagger\psi_\beta^\dagger\psi_\gamma\psi_\delta(\vec R)+...\
\end{align}
for some constant coefficients $C^{\sigma\sigma'}_{\alpha\beta\gamma\delta}$ which we would determine in the main text.

\section{Amplitudes}
In this section, we compute the $2\rightarrow 2'$ scattering amplitude. This will also provide us with an effective vertex that will be used for the later computations. It would be natural to compute the scattering process involving the dressed excitations since they are the eigenstates of the single-particle Hamiltonian. However, as it is obvious from the expression~\eqref{UnitaryTrans.1}, interactions between these excitations become involved and also $\vec {k} $- dependent. For this reason, we compute the scattering process between the original excitations and later diagonalize the scattering amplitude to obtain the corresponding results for the dressed excitations.

Let us first calculate the propagator. We start with the definition, 
\begin{align}
&G_{01}(E,\vec k)=\frac{i}{E-\frac{\vec k_\perp^2}{2}-\frac{1}{2}(k_x+\kappa_0)^2+i\epsilon},\\
&G_{02}(E,\vec k)=\frac{i}{E-\frac{\vec k_\perp^2}{2}-\frac{1}{2}(k_x-\kappa_0)^2+i\epsilon}\,.
\end{align}
The above is the propagator for $\psi_i$ fields in absence of $\Omega$. In the presence of $\Omega$, we have the following the propagator in the momentum-space,
\begin{align}
&G_{11}(E,\vec k)=\frac{G_{01}(E,\vec k)}{1+\frac{\Omega^2}{4}G_{01}(E,\vec k)G_{02}(E,\vec k)},\\
&G_{22}(E,\vec k)=\frac{G_{02}(E,\vec k)}{1+\frac{\Omega^2}{4}G_{01}(E,\vec k)G_{02}(E,\vec k)}\,,\\
&G_{12}(E,\vec k)=G_{21}(E,\vec k)=-\frac{i\frac{\Omega}{2}G_{02}(E,\vec k)G_{01}(E,\vec k)}{1+\frac{\Omega^2}{4}G_{01}(E,\vec k)G_{02}(E,\vec k)}\,.
\end{align}
In the matrix form, we write the Green's function as
\begin{equation}\label{GreensFn.1}
\mathcal G(E,\vec k)=\begin{pmatrix}G_{11}(E,\vec k)&G_{12}(E,\vec k)\\G_{21}(E,\vec k)&G_{22}(E,\vec k)\end{pmatrix}\,.
\end{equation}
The Green's function has poles at
\beq
E_\pm(\vec k)=\frac{1}{2}\vec k^2+\frac{1}{2}\kappa^2_0\pm\sqrt{\frac{\Omega^2}{4}+k_x^2\kappa_0^2}\,,
\eeq
which corresponds to the dispersion relation for the on-shell dressed excitations.
The inverse of the Green's function~\eqref{GreensFn.1} is
\beq
\mathcal G(E,\vec k)^{-1}=\begin{pmatrix}G_{01}^{-1}(E,\vec k)&\frac{i\Omega}{2}\\\frac{i\Omega}{2}&G_{02}^{-1}(E,\vec k)\end{pmatrix}\,.
\eeq
We will compute the amplitude in the centre of mass frame where the incoming particle with energy and momentum $(E_1,\vec p)$ and $(E_2,-\vec p)$ will scatter to outgoing particles with energy and momentum being $(E_3,\vec p')$ and $(E_4,-\vec p')$. At the moment, we will not assume that $E_i$'s satisfy the dispersion relation of the physical excitations. To compute the S-matrix, we follow the LSZ reduction formula which implies that the scattering amplitude for the dressed excitations is extracted by evaluating the residue of the 4-point Green's function at the poles where external states are on-shell. It is important to emphasise that in the Raman coupling case, i.e. $\kappa_0\neq0$, the Hamiltonian is not Galilean invariant, and as a result, the amplitude will not be Galilean invariant. The amplitude in the general frame will depend on the $x$-components of the external momentum. 

\begin{figure}[htpb]
\begin{center}
\vspace{1cm}
\includegraphics[width=3.5in]{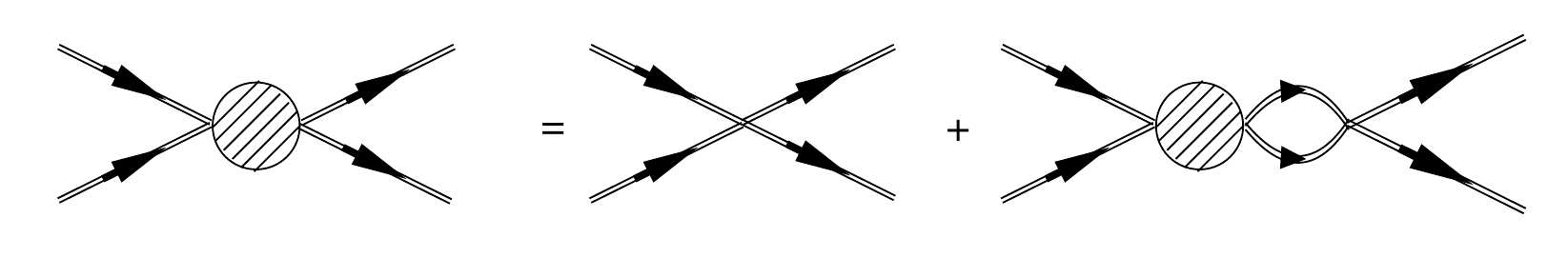}
\end{center}
\caption{The 2$\rightarrow$2 scattering diagram. The double line represents the matrix valued propagator.}\label{AmplitudeFig.}
\end{figure}

As mentioned above, we will focus on the scattering process in the centre-of-mass frame, in which case we will be interested in computing
\begin{align}
&\sum_{a,b,c,d}\mathcal G_{ia}(E_1,\vec p)\mathcal G_{jb}(E_2,-\vec p)\mathcal G_{ck}(E_3,\vec p')\mathcal G_{d\ell}(E_4,-\vec p')\times\nn\\
&\qquad\qquad\qquad\qquad\qquad\qquad\qquad\times \Big(i\mathcal M_{ab\rightarrow cd}(E)\Big)
\end{align}
with $E_1+E_2=E_3+E_4=E$. Here $a,b,..$ indices take value $1$ and $2$. Also, $\mathcal M_{ab\rightarrow cd}(E)$ is the amplitude for the incoming atoms in the spin states $a$ and $b$ scatter to outgoing spin states labelled by $c$ and $d$. The above computation can be represented in the diagrammatic form as shown in the figure~\ref{AmplitudeFig.}.

Using these Feynman diagrams, we can write down the amplitude as
\begin{align}\label{MatrixAmp.}
&i\mathcal M_{ab\rightarrow cd}(E)=-\frac{ig_{ef}}{2}(\delta_{ae}\delta_{bf}+\delta_{af}\delta_{be})(\delta_{ce}\delta_{df}+\delta_{cf}\delta_{de})\nn\\
&+(-\frac{ig_{ef}}{2})(\delta_{ae}\delta_{bf}+\delta_{af}\delta_{be})\times\nn\\
&\int\frac{ded^3q}{(2\pi)^{d+1}}G_{eu}(e,\vec q)G_{fv}(E-e,-\vec q)i\mathcal M_{uv\rightarrow cd}(E)\,.\nn\\
\end{align}
In the above, we have followed the Einstein's summation convention, i.e. the repeated indices are assumed to be summed over. The components of the coupling matrix are $g_{11}=g_{22}=g$ and $g_{12}=g_{21}=\lambda$.
Based on the one-loop computation, we make the following simplifying assumptions about the components of the matrix elements, 
\begin{align}
&\mathcal M_{11\rightarrow 11}=\mathcal M_{22\rightarrow 22}=\mathcal A(E)\,,\nn\\
&\mathcal M_{12\rightarrow 12}=\tilde{\mathcal A}(E)\,,\nn\\
&\mathcal M_{11\rightarrow 22}=\mathcal M_{22\rightarrow 11}=\hat{\mathcal A}(E)\,,\nn\\
&\mathcal M_{12\rightarrow 11}=\mathcal M_{12\rightarrow 22}=\mathcal M_{11\rightarrow 12}=\mathcal M_{22\rightarrow 12}={\mathcal A}'(E)\,.\nn\\
\end{align}
Then, from the eq.~\eqref{MatrixAmp.}, we can write the following matrix equation for the amplitudes\,,
\begin{widetext}
\begin{align}
\begin{pmatrix}i\mathcal A\\i\tilde {\mathcal  A}\\i\mathcal A'\\i\hat {\mathcal A}\end{pmatrix}=\begin{pmatrix}-2ig\\-i\lambda\\0\\0\end{pmatrix}-i\begin{pmatrix}g\mathcal Z_1&0&2g\mathcal Z_2&g\mathcal Z_3\\0&\lambda(\mathcal Z_3+\mathcal Z_4)&2\lambda \mathcal Z_2&0\\\lambda \mathcal Z_2&0&\lambda(\mathcal Z_3+\mathcal Z_4)&\lambda \mathcal Z_2\\g\mathcal Z_3&0&2g\mathcal Z_2&g\mathcal Z_1\end{pmatrix}\cdot\begin{pmatrix}i\mathcal A\\i\tilde {\mathcal A}\\i\mathcal A'\\i\hat {\mathcal A}\end{pmatrix}\,.
\end{align}
Solving the above, we can obtain the solutions which are
\begin{align}\label{ExplicitFormAmp}
&i\mathcal A=-\frac{2ig\Big[2g\lambda \mathcal Z_2^2+(1+ig \mathcal Z_1)(1+i\lambda(\mathcal Z_3+\mathcal Z_4))\Big]}{(1+ig(\mathcal Z_1-\mathcal Z_3))\Big[(1+ig(\mathcal Z_1+\mathcal Z_3))(1+i\lambda(\mathcal Z_3+\mathcal Z_4))+4g\lambda \mathcal Z_2^2\Big]}\,,\nn\\
&i\tilde {\mathcal A}=-\frac{i\lambda(1+ig(\mathcal Z_1+\mathcal Z_3))}{\Big[(1+ig(\mathcal Z_1+\mathcal Z_3))(1+i\lambda(\mathcal Z_3+\mathcal Z_4))+4g\lambda \mathcal Z_2^2\Big]}\,,\nn\\
&i\mathcal A'=-\frac{2g\lambda\mathcal Z_2}{\Big[(1+ig(\mathcal Z_1+\mathcal Z_3))(1+i\lambda(\mathcal Z_3+\mathcal Z_4))+4g\lambda \mathcal Z_2^2\Big]}\,,\nn\\
&i\hat {\mathcal A}=\frac{2ig^2(i\mathcal Z_3+2\lambda\mathcal Z^2_2-\lambda\mathcal Z_3(\mathcal Z_3+\mathcal Z_4))}{(1+ig(\mathcal Z_1-\mathcal Z_3))\Big[(1+ig(\mathcal Z_1+\mathcal Z_3))(1+i\lambda(\mathcal Z_3+\mathcal Z_4))+4g\lambda \mathcal Z_2^2\Big]}\,.\nn\\
\end{align}
\end{widetext}
Here, the functions $\mathcal Z_i(E)$ are one-loop integrals. For more details about these function, see the appendix~\ref{OneLoopCompt.}\,. After partially evaluating the one-loop integrals, the $\mathcal Z_i(E)$ can be organized in the form, 
\begin{align}\label{OneLoopSpinOrbit}
&\mathcal Z_1(E)=\frac{i}{4\pi}\sqrt{-E+\kappa_0^2-i\epsilon}+\frac{i\Omega^2}{2}I_1(E)\,,\nn\\
&\mathcal Z_2(E)=\frac{i\Omega}{2}I_2(E),\quad \mathcal Z_3(E)=\frac{i\Omega^2}{2}I_1(E)\,,\nn\\
&\mathcal Z_4(E)=\frac{i}{4\pi}\sqrt{-E+\kappa_0^2-i\epsilon}+4i\kappa_0^2 I_3(E)+i\frac{\Omega^2}{2}I_1(E)\,.\nn\\
\end{align}

Now, at this important, let us emphasize the importance of working in the variables $\psi$, rather than in $\alpha$. As mentioned before, the physical excitations are dressed particles; therefore, it is natural to discuss the scattering amplitude of these excitations. These amplitudes can be easily obtained from the above and are given as the linear combination of the amplitudes $\mathcal M_{ab\rightarrow cd}(E)$ as we will discuss below. The point is this that since the unitary transformation~\eqref{UnitaryTrans.1} depends on the $x$-component of the momentum, $k_x$, as a result, the interaction terms are also dependent on $k_x$. This makes the scattering amplitude momentum dependent for dressed excitations. This dependence, however, disappear for $\kappa_0=0$.

Using the LSZ reduction formula, the amplitude for the dressed excitation is
\begin{align}
&i\mathcal V_{ij\rightarrow k\ell}=\prod_{i=1}^4(E_i-E_{\delta_i}(k_i))U^{-1}_{mi}U^{-1}_{nj}U_{kp}U_{\ell q}\times\nn\\
&\times\mathcal G_{ma}(E_1,\vec p)\mathcal G_{nb}(E_2,-\vec p)\mathcal G_{cp}(E_3,\vec p')\mathcal G_{dq}(E_4,-\vec p')i\mathcal M_{ab\rightarrow cd}(E)\,.\nn\\
\end{align}
The above needs to be evaluated for the on-shell incoming/outgoing particles with energy $E=E_+(p)$ and/or $E=E_-(p)$, where the matrix $U$ is given in~\eqref{UnitaryTrans.1}. For example, the amplitude for the dressed excitation $(+,+)\rightarrow (+,+)$ with ingoing and outgoing kinematical variables $(E_+(\vec k),E_+(-\vec k))$ and $(E_+(\vec {\tilde k}),E_+(-\vec {\tilde k}))$, respectively is
\begin{align}\label{PhysicalExctAmp}
&i\mathcal V_{++\rightarrow ++}=i(\mathcal A+\hat{\mathcal A})\frac{\Omega^2}{2\tilde\Omega\tilde\Omega'}+i\tilde{\mathcal A}+\frac{\Omega}{\tilde\Omega\tilde\Omega'}(\tilde\Omega+\tilde\Omega')i\mathcal A'\,.
\end{align}
Here $\tilde\Omega=\sqrt{\Omega^2+4\kappa_0^2k_x^2}$ and $\tilde\Omega'=\sqrt{\Omega^2+4\kappa_0^2\tilde{k}_x^2}$\,. Note that the amplitude for the dressed particles depends on the $x$-component of the momentum. Furthermore, using the explicit form of the amplitude in~\eqref{ExplicitFormAmp}, we see that the expression~\eqref{PhysicalExctAmp} has poles at
\beq
(1+ig(\mathcal Z_1+\mathcal Z_3))(1+i\lambda(\mathcal Z_3+\mathcal Z_4))+4g\lambda \mathcal Z_2^2=0\,.
\eeq
This is the generalization of the typical poles encountered in non-relativistic scattering amplitude.
\section{Coherently Coupled system}
In this section, we consider the coherently coupled bosonic system. It corresponds to setting the Raman coupling $\kappa_0=0$. 
The physical excitations correspond to the dispersion relation
\beq
E_\pm(\vec k)=\frac{k^2}{2}\pm\frac{\Omega}{2}\,.
\eeq
In this case, the integrals appearing in the one loop computations~\eqref{OneLoopSpinOrbit} simplifies considerably and are given as
\begin{align}
&Z_1(E)=\frac{i}{4\pi}\sqrt{-E-i\epsilon}+\frac{1}{16\pi}\Big[-2\sqrt{E+i\epsilon}\nn\\
&\qquad\qquad\qquad\qquad\quad+\sqrt{E+i\epsilon+\Omega}+\sqrt{E+i\epsilon-\Omega}\Big]\,,\nn\\
&Z_2(E)=-\frac{1}{16\pi}(\sqrt{E+i\epsilon+\Omega}-\sqrt{E+i\epsilon-\Omega})\,,\nn\\
&Z_3(E)=\frac{1}{16\pi}\Big[\sqrt{E+i\epsilon+\Omega}+\sqrt{E+i\epsilon-\Omega}-2\sqrt{E+i\epsilon}\Big]\,,\nn\\
&Z_4(E)=\frac{i}{4\pi}\sqrt{-E-i\epsilon}+\frac{1}{16\pi}\Big[-2\sqrt{E+i\epsilon}\,,\nn\\
&\qquad\qquad\qquad\qquad+\sqrt{E+i\epsilon+\Omega}+\sqrt{E+i\epsilon-\Omega}\Big]\,.
\end{align}
The expression for the S-matrix for the dressed particles also simplifies. For example, the amplitude for $(+,+)\rightarrow(+,+)$ is
\begin{align}\label{PhysicalExctAmp2}
&i\tilde{\mathcal V}_{++\rightarrow ++}=\frac{i}{2}(\mathcal A+\hat{\mathcal A}+2\tilde{\mathcal A})+2i\mathcal A'\,.
\end{align}
In this case, the location of poles in the S-matrix is determined by solutions of the equation,
\beq
(1+ig(\mathcal Z_1+\mathcal Z_3))(1+i\lambda(\mathcal Z_3+\mathcal Z_1))+4g\lambda \mathcal Z_2^2=0\,.
\eeq
Next, we compute the OPE of the operators $\psi^\dagger_\sigma(\vec R-\frac{1}{2}\vec r)$ and $\psi_{\sigma'}(\vec R+\frac{1}{2}\vec r)$. The OPE as an operator statement must be true when inserted among the chain of operators while computing a correlation function or expectation value. Following the analysis presented in~\cite{braaten2010short}, we compute the OPE by evaluating the expectation value of the bilocal operator $<\psi^\dagger_\sigma(\vec R-\frac{1}{2}\vec r)\psi_{\sigma'}(\vec R+\frac{1}{2}\vec r)>$ in two-particle states. The computation of the expectation value can be given a Feynman diagram representation, in which case it is a sum of three types of diagrams: one involves no scattering, the second involves single scattering, and the third involves two scattering. Each of these diagrams contributes to the small $\vec r$ expansion of the OPE. Restricting ourselves to the first leading order in $r=|\vec r|$, there are contributions that are analytic in $\vec r$ and others that are non-analytic, i.e., depending on $r$ rather than on $\vec r$. Our interests are in the terms whose Wilson coefficients are non-analytic in $\vec r$. It is not very difficult to guess the source of these contributions. It is the Feynman diagram which involves the integration over the momentum dual to the $\vec r$ that will give the contribution non-analytic in $\vec r$.
In fact, all other diagrams, with no and single scattering, where the momentum dual to $\vec r$ is fixed by the conservation of energy and momentum, are analytic in $\vec r$. 
For example, the two diagrams with one scattering contribute in the 4-point correlation function as
\begin{widetext}
\begin{align}
&G_{ia}(E_1,\vec p)G_{jb}(E_2,-\vec p)\Big(i\mathcal M_{ab\rightarrow cd}\Big)\Big[G_{c\sigma'}(E_2',-\vec p')G_{dk}(E_1',\vec p') G_{\sigma\ell}(E_2',-\vec p')e^{i\vec p'\cdot\vec r}+G_{c\sigma'}(E_1',\vec p')G_{d\ell}(E_2',-\vec p')\times\nn\\
&\times G_{\sigma k}(E_1',\vec p')e^{-i\vec p'\cdot\vec r}\Big]+\Big[G_{ja}(E_2,-\vec p)G_{\sigma b}(E_1,\vec p)G_{i\sigma'}(E_1,\vec p)e^{i\vec p\cdot\vec r}+G_{ia}(E_1,\vec p)G_{\sigma b}(E_2,-\vec p)G_{j\sigma'}(E_2,-\vec p)e^{-i\vec p\cdot\vec r}\Big]\times\nn\\
&\qquad\qquad\qquad\qquad\qquad\qquad\qquad\qquad\qquad\qquad\qquad\qquad\times\Big(i\mathcal M_{ab\rightarrow cd}\Big)G_{cp}(E_1',\vec p')G_{dq}(E_2',-\vec p')\,.
\end{align}
\end{widetext}
Note that these contributions are analytic in $\vec r$, i.e. doing the Taylor series expansion, we obtain terms depending on $\vec r$ only.

\begin{figure}[htpb]
\begin{center}
\includegraphics[width=3.5in]{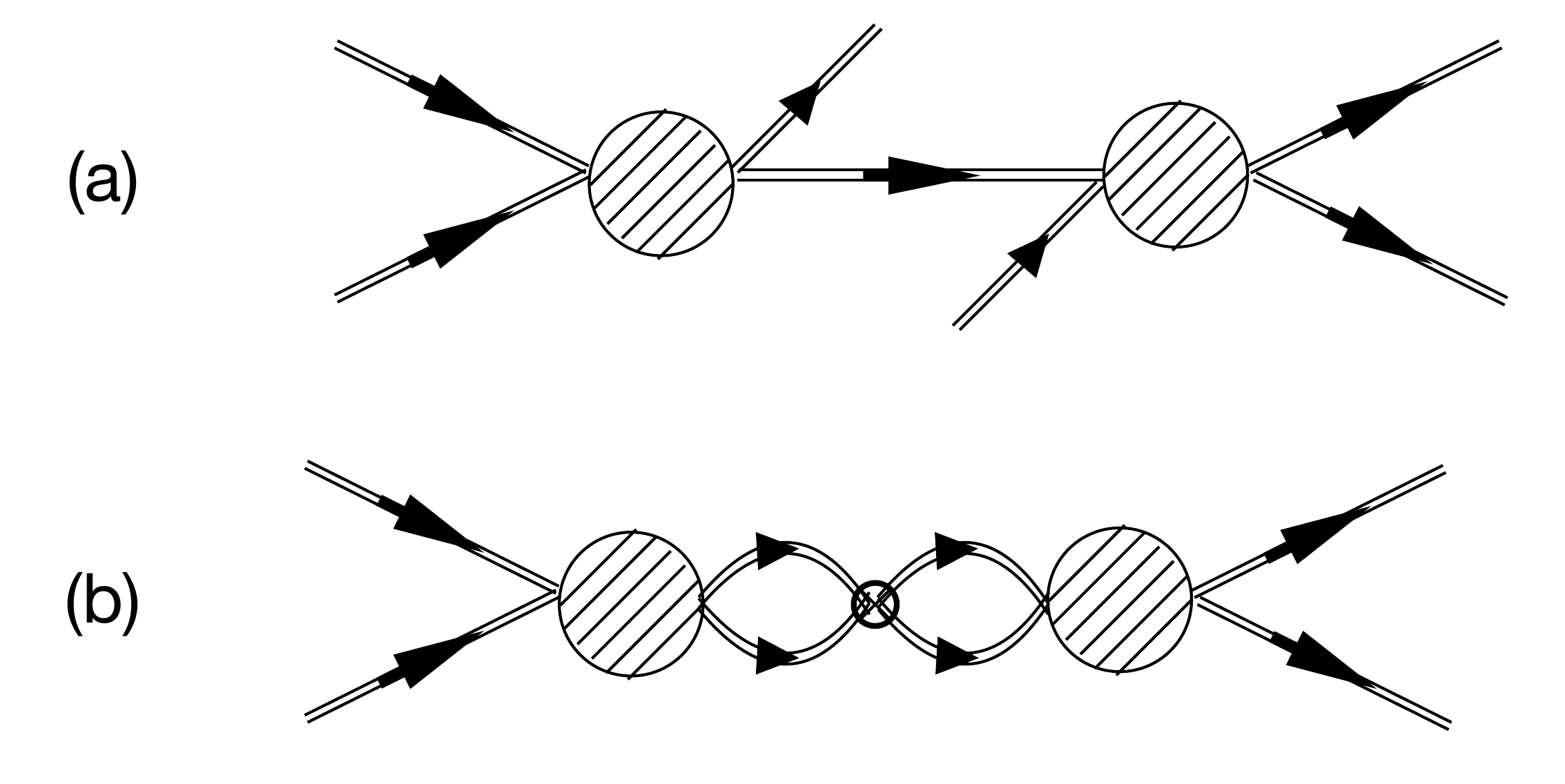}
\end{center}
\caption{The figure in (a) computes the expectation value of the operator $\psi^\dagger_\sigma(\vec r)$ and $\psi^\dagger_{\sigma'}(\vec r')$. The figure in (b) computes the expectation value of the local operator $\psi^\dagger_a\psi^\dagger_b\psi_c\psi_d(\vec r)$. }\label{ExpValueofBilocalOp}
\end{figure}

Next, we compute the diagrams with two scattering. The diagram is shown in the figure (a) of~\ref{ExpValueofBilocalOp}. These diagrams with two scattering contribute to the OPE with terms that may not be analytic in $\vec {r} $. For the $2\rightarrow 2'$ scattering process, i.e. $i,j\rightarrow k,\ell$, and with operators $\psi_{\sigma'}(\vec R+\frac{1}{2}\vec r)$ and $\psi^\dagger_\sigma(\vec R-\frac{1}{2}\vec r)$, inserted, we obtain  
\begin{align}\label{MatrixelementOPE}
&-\int\frac{ded^3p}{(2\pi)^4}e^{-i\vec p\cdot\vec r}\Big[\mathcal M_{ij\rightarrow ab}\mathcal M_{cd\rightarrow k\ell}G_{a\sigma'}(e,\vec p)G_{bc}(E-e,-\vec p)\times\nn\\
&\qquad\qquad\qquad\qquad\qquad\qquad\qquad\times G_{\sigma d}(e,\vec p)\Big]\,.
\end{align}
Here $i,j,k,\ell$ are external indices. For example, if we evaluate the above for  $\sigma=\sigma'=1$, then we have the expectation value 
\begin{widetext}
\begin{align}\label{MatrixelementOPE.0}
&-\int\frac{ded^3p}{(2\pi)^4}e^{-i\vec p\cdot\vec r}\Big[\mathcal M_{ij\rightarrow 11}\mathcal M_{11\rightarrow k\ell}G_{11}(e,\vec p)^2G_{11}(E-e,-\vec p)+\mathcal M_{ij\rightarrow 11}\mathcal M_{12\rightarrow k\ell}G_{11}(e,\vec p)\Big(G_{12}(e,\vec p)G_{11}(E-e,-\vec p)\nn\\
&+G_{11}(e,\vec p)G_{12}(E-e,-\vec p)\Big)+\mathcal M_{ij\rightarrow 11}\mathcal M_{22\rightarrow k\ell}G_{11}(e,\vec p)G_{12}(e,\vec p)G_{12}(E-e,-\vec p)\nn\\
&+\mathcal M_{ij\rightarrow 12}\mathcal M_{11\rightarrow k\ell}G_{11}(e,\vec p)\Big(G_{21}(e,\vec p)G_{11}(E-e,-\vec p)+G_{11}(e,\vec p)G_{12}(E-e,-\vec p)\Big)+\nn\\
&+\mathcal M_{ij\rightarrow 12}\mathcal M_{12\rightarrow k\ell}\Big[G_{21}(e,\vec p)\Big(G_{12}(e,\vec p)G_{11}(E-e,-\vec p)+G_{12}(E-e,-\vec p)G_{11}(e,\vec p)\Big)+G_{11}(e,\vec p)\Big(G_{12}(e,\vec p)G_{12}(E-e,-\vec p)\nn\\
&+G_{22}(E-e,-\vec p)G_{11}(e,\vec p)\Big)\Big]+\mathcal M_{ij\rightarrow 12}\mathcal M_{22\rightarrow k\ell}\Big(G_{12}(e,\vec p)^2G_{12}(E-e,-\vec p)+G_{12}(e,\vec p)G_{11}(e,\vec p)G_{22}(E-e,-\vec p)\Big)\nn\\
&+\mathcal M_{ij\rightarrow 22}\mathcal M_{11\rightarrow k\ell}G_{12}(e,\vec p)G_{11}(e,\vec p)G_{12}(E-e,-\vec p)+\mathcal M_{ij\rightarrow 22}\mathcal M_{12\rightarrow k\ell}G_{12}(e,\vec p)\Big(G_{12}(e,\vec p)G_{12}(E-e,-\vec p)\nn\\
&+G_{11}(e,\vec p)G_{22}(E-e,-\vec p)\Big)+\mathcal M_{ij\rightarrow 22}\mathcal M_{22\rightarrow k\ell}G_{12}(e,\vec p)^2G_{22}(E-e,-\vec p)\Big]\,.
\end{align}
\end{widetext}
The above integrals contribute terms in OPE that are not analytic in $\vec r$. The expression for the various integrals, given in terms of Fourier transform of the product of Green's function, appearing above are given in the appendix~\ref{ProdExpGreensFn}. We have evaluated each of the above integrals and expanded in powers of $r$, see the appendix~\ref{ProdGreensFnCohenertlyCoupled}. Thus, if we are looking at the leading non-analytic terms i.e. the contribution proportional to $r$, then we get
\begin{align}\label{MatrixelementOPE.1}
&-\int\frac{ded^3p}{(2\pi)^4}e^{-i\vec p\cdot\vec r}G_{11}(e,\vec p)^2\Big[\mathcal M_{ij\rightarrow 11}\mathcal M_{11\rightarrow k\ell}G_{11}(E-e,-\vec p)\nn\\
&\qquad\qquad+\mathcal M_{ij\rightarrow 12}\mathcal M_{12\rightarrow k\ell}G_{22}(E-e,-\vec p)\Big]\,,\nn\\
&\stackrel{\mathcal O(r)}{=}-\frac{r}{8\pi}\Big[\mathcal M_{ij\rightarrow 11}\mathcal M_{11\rightarrow k\ell}+\mathcal M_{ij\rightarrow 12}\mathcal M_{12\rightarrow k\ell}\Big]\,.
\end{align}

Now, we want to find the local operator whose expectation value in the two-particle state gives the above contribution. It will give us the form of the contact density operator. Looking at the expression in~\eqref{MatrixelementOPE.1}, it is not difficult to guess the form of the contact density operator. It turns out that for the diagonal case, i.e. $\sigma=\sigma'$, the operator is a linear combination of operators of the form $\mu_1(\psi^\dagger_1\psi_1)^2+\mu_2(\psi^\dagger_2\psi_2)^2+\mu_3\psi^\dagger_1\psi_1\psi^\dagger_2\psi_2$. We can compute the expectation value for this linear combination between $i,j\rightarrow k,\ell$ states. In the Feynman diagram computation, we need to evaluate the digram (b) in the figure~\ref{ExpValueofBilocalOp}. Results of these computations are given in the appendix~\ref{AppendixExpectationaValueOPE}.
Comparing the expressions~\eqref{MatrixelementOPE.1} and~\eqref{ExpectationaValueOPE.1}, we can fix the coefficients $\m_i$.
For example, if we evaluate~\eqref{ExpectationaValueOPE.1} for $i=k=1, j=\ell=2$, and look for only terms giving $r$, then we get
\begin{align}\label{ExpBilocal1}
&\m_3+2\m_3\frac{i\tilde {\mathcal A}+i\lambda}{-i\lambda}+\m_3(\frac{i\tilde {\mathcal A}+i\lambda}{-i\lambda})^2+\m_1(\frac{i\mathcal A'}{-ig})^2+\m_2\frac{\mathcal A^{'2}}{g^2}\nn\\
&=\frac{\m_3}{\lambda^2}\tilde {\mathcal A}^2+\frac{\m_1+\m_2}{g^2}{\mathcal A}^{'2}\,,
\end{align}
and repeating the above calculation for $i=j=1$ and $k=1,\ell=2$, we obtain
\begin{align}\label{ExpBilocal2}
&2\m_1(\frac{i{\mathcal A}'}{-ig})+\m_3(\frac{i{\mathcal A}'}{-i\lambda})+\m_1(\frac{i\mathcal A+2ig}{-ig})(\frac{i{\mathcal A}'}{-ig})\nn\\
&+\m_3(\frac{i{\mathcal A}'}{-i\lambda})(\frac{i\tilde {\mathcal A}+i\lambda}{-i\lambda})+\m_2(\frac{i\hat{\mathcal A}}{-ig})(\frac{i{\mathcal A}'}{-ig})=\frac{\m_1}{g^2}{\mathcal A}{\mathcal A}'+\frac{\m_3}{\lambda^2}{\mathcal A}'\tilde {\mathcal A}\nn\\
&+\frac{\m_2}{g^2}\hat{\mathcal A}{\mathcal A}'\,.
\end{align}
Comparing the expressions in~\eqref{ExpBilocal1} and~\eqref{ExpBilocal2} with~\eqref{MatrixelementOPE.1}, we find that $\m_1=\frac{g^2}{8\pi}$, $\m_2=0$, and $\m_3=\frac{\lambda^2}{8\pi}$.. 
Thus, we get the OPE
\begin{align}
&\psi^\dagger_1(\vec R-\frac{1}{2}\vec r)\psi_1(\vec R+\frac{1}{2}\vec r)=\psi^\dagger_1\psi_1(\vec R)+\vec r\cdot\psi^\dagger_1\overset{\leftrightarrow}{\nabla}\psi_1(\vec R)\nn\\
&\qquad-\frac{r}{8\pi}\Big[g^2(\psi^\dagger_1\psi_1)^2+\lambda^2\psi^\dagger_1\psi_1\psi^\dagger_2\psi_2\Big](\vec R)+..
\end{align}
Here $\psi^\dagger_1\overset{\leftrightarrow}{\nabla}\psi_1=\frac{1}{2}(\psi^\dagger_1\nabla\psi_1-\nabla\psi^\dagger_1\psi_1)$.

Similarly, we can compute the OPE for the off-diagonal operators in the spin space. We repeat the above calculation for $\sigma'=1$ and $\sigma=2$, i.e. between $\psi_{1}$ and $\psi^\dagger_2$, we obtain
\begin{widetext}
\begin{align}
&-\int\frac{ded^3p}{(2\pi)^4}e^{-i\vec p\cdot\vec r}\Big[\mathcal M_{ij\rightarrow 11}\mathcal M_{11\rightarrow pq}G_{11}(e,\vec p)G_{12}(e,\vec p)G_{11}(E-e,-\vec p)+\mathcal M_{ij\rightarrow 11}\mathcal M_{12\rightarrow pq}G_{11}(e,\vec p)\Big(G_{22}(e,\vec p)G_{11}(E-e,-\vec p)\nn\\
&+G_{21}(e,\vec p)G_{12}(E-e,-\vec p)\Big)+\mathcal M_{ij\rightarrow 11}\mathcal M_{22\rightarrow pq}G_{11}(e,\vec p)G_{22}(e,\vec p)G_{12}(E-e,-\vec p)\nn\\
&+\mathcal M_{ij\rightarrow 12}\mathcal M_{12\rightarrow pq}\Big[G_{11}(e,\vec p)\Big(G_{22}(e,\vec p)G_{21}(E-e,-\vec p)+G_{21}(e,\vec p)G_{22}(E-e,-\vec p)\Big)+
\nn\\
&+G_{12}(e,\vec p)G_{22}(e,\vec p)G_{11}(E-e,-\vec p)+G_{12}(e,\vec p)^2G_{12}(E-e,-\vec p)\Big]\nn\\
&+\mathcal M_{ij\rightarrow 12}\mathcal M_{11\rightarrow pq}\Big[G_{11}(e,\vec p)G_{12}(e,\vec p)G_{12}(E-e,-\vec p)+G_{12}(e,\vec p)^2G_{11}(E-e,-\vec p)\Big]\nn\\
&+\mathcal M_{ij\rightarrow 12}\mathcal M_{22\rightarrow pq}\Big(G_{11}(e,\vec p)G_{22}(e,\vec p)G_{22}(E-e,-\vec p)+G_{12}(e,\vec p)G_{22}(e,\vec p)G_{12}(E-e,-\vec p)\Big)\nn\\
&+\mathcal M_{ij\rightarrow 22}\mathcal M_{11\rightarrow pq}G_{12}(e,\vec p)^2G_{12}(E-e,-\vec p)+\mathcal M_{ij\rightarrow 22}\mathcal M_{12\rightarrow pq}G_{12}(e,\vec p)\Big(G_{22}(e,\vec p)G_{12}(E-e,-\vec p)\nn\\
&+G_{12}(e,\vec p)G_{22}(E-e,-\vec p)\Big)+\mathcal M_{ij\rightarrow 22}\mathcal M_{22\rightarrow pq}G_{12}(e,\vec p)G_{22}(E-e,-\vec p)G_{22}(e,\vec p)\Big]
\end{align}
\end{widetext}
From the above we can collect the terms giving terms linear in $r$. These are
\beq
-\frac{r}{8\pi}\Big[\mathcal M_{ij\rightarrow 11}\mathcal M_{12\rightarrow pq}+\mathcal M_{ij\rightarrow 12}\mathcal M_{22\rightarrow pq}\Big]\,.
\eeq
An important point to note that the coefficient of the above expression does not depend on $\Omega$.
Now, we compute the expectation value of the chain of operators of the form $\eta_1\psi^\dagger_2\psi_1\psi^\dagger_1\psi_1+\eta_2\psi^\dagger_2\psi_1\psi^\dagger_2\psi_2+\eta_3\psi^\dagger_1\psi_2\psi^\dagger_1\psi_1+\eta_4\psi^\dagger_1\psi_2\psi^\dagger_2\psi_2$. For $i,j\rightarrow k,\ell$ states, we obtain~\ref{ExpectationaValueOPE.2}.
Thus, repeating the same exercise as in the diagonal case, we obtain the OPE to be
\begin{align}
&\psi^\dagger_2(\vec R-\frac{1}{2}\vec r)\psi_1(\vec R+\frac{1}{2}\vec r)=\psi^\dagger_2\psi_1(\vec R)+\vec r\cdot\psi^\dagger_2\overset{\leftrightarrow}{\nabla}\psi_1\nn\\
&\qquad\qquad-\frac{r}{8\pi}\lambda g\Big[\psi^\dagger_2\psi_1\psi^\dagger_1\psi_1+\psi^\dagger_2\psi_1\psi^\dagger_2\psi_2\Big]+..
\end{align}
We can repeat the similar steps for the other values of $\sigma$ and $\sigma'$. We list here all the OPE to the leading order in $r$ (non-analytic):
\begin{align}
&1.\,\,\psi^\dagger_1(\vec R-\frac{1}{2}\vec r)\psi_1(\vec R+\frac{1}{2}\vec r)=\psi^\dagger_1\psi_1(\vec R)+\vec r\cdot\psi^\dagger_1\overset{\leftrightarrow}{\nabla}\psi_1(\vec R)\nn\\
&-\frac{r}{8\pi}\Big[g^2(\psi^\dagger_1\psi_1)^2+\lambda^2\psi^\dagger_1\psi_1\psi^\dagger_2\psi_2\Big](\vec R)+...\,,\nn\\
&2.\,\,\psi^\dagger_2(\vec R-\frac{1}{2}\vec r)\psi_2(\vec R+\frac{1}{2}\vec r)=\psi^\dagger_2\psi_2(\vec R)+\vec r\cdot\psi^\dagger_2\overset{\leftrightarrow}{\nabla}\psi_2\nn\\
&-\frac{r}{8\pi}\Big[g^2(\psi^\dagger_2\psi_2)^2+\lambda^2\psi^\dagger_1\psi_1\psi^\dagger_2\psi_2](\vec R)+...\,,\nn\\
&3.\,\,\psi^\dagger_2(\vec R-\frac{1}{2}\vec r)\psi_1(\vec R+\frac{1}{2}\vec r)=\psi^\dagger_2\psi_1(\vec R)+\vec r\cdot\psi^\dagger_2\overset{\leftrightarrow}{\nabla}\psi_1\nn\\
&-\frac{r}{8\pi}\lambda g\Big[\psi^\dagger_2\psi_1\psi^\dagger_1\psi_1+\psi^\dagger_2\psi_1\psi^\dagger_2\psi_2\Big](\vec R)+...\,,\nn\\
&4.\,\,\psi^\dagger_1(\vec R-\frac{1}{2}\vec r)\psi_2(\vec R+\frac{1}{2}\vec r)=\psi^\dagger_1\psi_2(\vec R)+\vec r\cdot\psi^\dagger_1\overset{\leftrightarrow}{\nabla}\psi_2\nn\\
&-\frac{r}{8\pi}\lambda g\Big[\psi^\dagger_1\psi_2\psi^\dagger_1\psi_1+\psi^\dagger_1\psi_2\psi^\dagger_2\psi_2\Big](\vec R)+...\,.
\end{align}
The above set of OPE is the main result of the paper. 

\section{Spin-orbit coupled case}
Now, we discuss the OPE in the spin-orbit coupled case. As we have emphasized before, we will focus only on the contribution from the 2-body physics, i.e. we ignore the contribution due to the Efimov effect. We look for the matrix element of the operators $\psi_\alpha^\dagger(\vec R-\frac{1}{2}\vec r)$ and $\psi_\beta^\dagger(\vec R+\frac{1}{2}\vec r)$. The computation proceeds in the same manner as we did in the case of the Rabi-coupled bosonic system. The expression for the matrix element between the states $i,j\rightarrow k,\ell$ remains the same as~\eqref{MatrixelementOPE}. The only difference comes from the explicit breaking of spatial isotropy due to the presence of the recoil momentum $\kappa_0$ along the $x$-direction. However, it is possible to argue that the OPEs, till the first order and non-analytic in $\vec r$, remain unchanged even in the presence of the recoil momentum. To see this, let us first understand the reason for the appearance of the non-analytic term in $\vec r$ in the previous integral. We have noticed that the source of the non-analytic terms is the integral of the form
\beq
\int\frac{d^3p}{(2\pi)^3}\frac{e^{i\vec p\cdot\vec r}}{(A-p^2)^2}\,.
\eeq
The above is not analytic in $\vec r$ at $\vec r=0$. Naively, if we differentiate the above with respect to $\vec r$ about $\vec r=0$, the integral vanishes. However, this conclusion is too quick, since the resulting integral is logarithmically divergent in the UV. This is the reason for the presence of the non-analytic term. On the other hand, if we consider the integral of the form
\beq
\int\frac{d^3p}{(2\pi)^3}\frac{e^{i\vec p\cdot\vec r}}{(A-p^2)^2\Big((A-p^2)^2-\Omega^2-4p_x^2\kappa_0^2\Big)}\,,
\eeq
then the above integral is analytic in $\vec r$ since the derivative exists (the resultant integral after derivative with respect to $\vec r$ is convergent). A similar conclusion can be drawn for the rest of the integrals appearing in the spin-orbit coupled case. For example, the integral of the form
\beq
\int\frac{d^3p}{(2\pi)^3}\frac{e^{i\vec p\cdot\vec r}p_x}{(A-p^2)\Big((A-p^2)^2-\Omega^2-4p_x^2\kappa_0^2\Big)^2}\,,
\eeq
will give terms analytic in $\vec r$. It is important to re-emphasize that we are only looking for the contributions in the OPE whose coefficients are non-analytic in $\vec r$. Clearly, there are terms in the OPE whose Wilson coefficients are analytic functions of $\vec r$ which we are not quantities of interest in this article.\\
Thus, we conclude that the OPEs of the field operators, till the first order and non-analytic in $\vec r$, in the spin-orbit coupled bosonic system are
\begin{align}
&1.\,\,\psi^\dagger_1(\vec R-\frac{1}{2}\vec r)\psi_1(\vec R+\frac{1}{2}\vec r)=\psi^\dagger_1\psi_1(\vec R)+\vec r\cdot\psi^\dagger_1\overset{\leftrightarrow}{\nabla}\psi_1(\vec R)\nn\\
&-\frac{r}{8\pi}\Big[g^2(\psi^\dagger_1\psi_1)^2+\lambda^2\psi^\dagger_1\psi_1\psi^\dagger_2\psi_2\Big](\vec R)+...\,,\nn\\
&2.\,\,\psi^\dagger_2(\vec R-\frac{1}{2}\vec r)\psi_2(\vec R+\frac{1}{2}\vec r)=\psi^\dagger_2\psi_2(\vec R)+\vec r\cdot\psi^\dagger_2\overset{\leftrightarrow}{\nabla}\psi_2\nn\\
&-\frac{r}{8\pi}\Big[g^2(\psi^\dagger_2\psi_2)^2+\lambda^2\psi^\dagger_1\psi_1\psi^\dagger_2\psi_2](\vec R)+...\,,\nn\\
&3.\,\,\psi^\dagger_2(\vec R-\frac{1}{2}\vec r)\psi_1(\vec R+\frac{1}{2}\vec r)=\psi^\dagger_2\psi_1(\vec R)+\vec r\cdot\psi^\dagger_2\overset{\leftrightarrow}{\nabla}\psi_1\nn\\
&-\frac{r}{8\pi}\lambda g\Big[\psi^\dagger_2\psi_1\psi^\dagger_1\psi_1+\psi^\dagger_2\psi_1\psi^\dagger_2\psi_2\Big](\vec R)+...\,,\nn\\
&4.\,\,\psi^\dagger_1(\vec R-\frac{1}{2}\vec r)\psi_2(\vec R+\frac{1}{2}\vec r)=\psi^\dagger_1\psi_2(\vec R)+\vec r\cdot\psi^\dagger_1\overset{\leftrightarrow}{\nabla}\psi_2\nn\\
&-\frac{r}{8\pi}\lambda g\Big[\psi^\dagger_1\psi_2\psi^\dagger_1\psi_1+\psi^\dagger_1\psi_2\psi^\dagger_2\psi_2\Big](\vec R)+...\,.
\end{align}

Next, we would like to see how the contact densities vary over the various phases of the spin-orbit coupled bosonic system. In the variational approach to finding the ground state of the system, we begin with the ansatz given by the coherent wave function, 
\beq
\begin{pmatrix}\psi_1\\\psi_2\end{pmatrix}=\sqrt{n}\Big[C_1\begin{pmatrix}\cos\theta\\-\sin\theta\end{pmatrix}e^{ik_1x}+C_2\begin{pmatrix}\sin\theta\\-\cos\theta\end{pmatrix}e^{-i\kappa_1x}e^{i\delta}\Big]\,,
\eeq
where $\cos2\theta=\frac{\kappa_1}{\kappa_0}$, $\delta$ is the phase difference between $C_1$ and $C_2$ and the constants $C_1$ and $C_2$ are determined by the normalization condition
\beq
\int d^3x\,\Big(|\psi_1|^2+|\psi_2|^2\Big)=N\,.
\eeq 
Here $N$ is the total number of atoms. There are three distinct phases separated by either a first-order or a continuous phase transition. These phases are: Plane wave phase, Zero momentum phase, and Stripe phase. These phases are realized as we vary Rabi and Raman coupling, $\Omega$ and $\kappa_0$, respectively.
We will evaluate the contact densities in each phase of the condensate. \\
{\bf Plane wave phase:} The plane wave phase is realized with $\kappa_1=\kappa_0\sqrt{1-\frac{\Omega^2}{4(\kappa_0^2-2G_2)^2}}$ with either $C_1=0$ or $C_2=0$. In this phase, the contact densities evaluate to
\begin{align}\label{ContactDen.PlaneWave}
&(\psi^\dagger_1\psi_1)^2(\vec R)=\frac{n^2}{4}\Big(1+\sqrt{1-\frac{\Omega^2}{4(\kappa_0^2-2G_2)^2}}\Big)^2\,, \nn\\
&(\psi^\dagger_2\psi_2)^2(\vec R)=\frac{n^2}{4}\Big(1-\sqrt{1-\frac{\Omega^2}{4(\kappa_0^2-2G_2)^2}}\Big)^2\,,\nn\\
&(\psi^\dagger_1\psi_1\psi^\dagger_2\psi_2)(\vec R)=n^2\frac{\Omega^2}{16(\kappa_0^2-2G_2)^2}\,,\nn\\
&(\psi^\dagger_2\psi_1\psi^\dagger_1\psi_1)(\vec R)=-\frac{n^2}{8}\frac{\Omega}{|\kappa_0^2-2G_2|}\Big[1+\sqrt{1-\frac{\Omega^2}{4(\kappa_0^2-2G_2)^2}}\Big]\,,\nn\\
&(\psi^\dagger_2\psi_1\psi^\dagger_2\psi_2)(\vec R)=-\frac{n^2}{8}\frac{\Omega}{|\kappa_0^2-2G_2|}\Big[1-\sqrt{1-\frac{\Omega^2}{4(\kappa_0^2-2G_2)^2}}\Big]\,.
\end{align}
Here $G_1=\frac{g+\lambda}{4}$ and $G_2=\frac{g-\lambda}{4}$.

\noindent{\bf Zero momentum phase:} The zero momentum phase is characterised by the constant value of the condensate. In this case, $\kappa_1=0$ and the net magnetization along the $z$-direction vanishes. In this case, the contact densities simplify to
\begin{align}\label{ContactDen.ZeroMom}
&(\psi^\dagger_1\psi_1)^2(\vec R)=(\psi^\dagger_2\psi_2)^2(\vec R)=\frac{n^2}{4}\,, \nn\\
&(\psi^\dagger_2\psi_1\psi^\dagger_1\psi_1)(\vec R)=(\psi^\dagger_1\psi_2\psi^\dagger_2\psi_2)(\vec R)=-\frac{n^2}{4}\,.
\end{align}
Note that the contact densities in the plane wave phase~\eqref{ContactDen.PlaneWave} continuously connect to their values in the zero momentum phase~\eqref{ContactDen.ZeroMom} and are continuous the phase boundary $\Omega=2(\kappa_0^2-2G_2)$. 

\noindent{\bf Stripe phase:} Next, we evaluate the contact densities in the stripe phase. This phase is characterised by the fact that it is a superposition of plane waves moving in the opposite direction along the $x$-axis. This gives rise to the density modulation. In this phase, we have $C_{1}=C_2$, $\kappa_1\neq0$ and magnetization along the $z$-axis vanishes. In this phase, we have
\begin{align}
&(\psi^\dagger_1\psi_1)^2(\vec R)=\frac{n^2}{4}\Big(1+\frac{\Omega}{2(G_1+\kappa_0^2)}\cos(2\kappa_1x-\delta)\Big)^2\,,\nn\\
&(\psi^\dagger_2\psi_2)^2(\vec R)=\frac{n^2}{4}\Big(1+\frac{\Omega}{2(G_1+\kappa_0^2)}\cos(2\kappa_1x-\delta)\Big)^2\,,\nn\\
&\psi^\dagger_1\psi_1\psi^\dagger_2\psi_2(\vec R)=\frac{n^2}{4}\Big(1+\frac{\Omega}{2(G_1+\kappa_0^2)}\cos(2\kappa_1x-\delta)\Big)^2\,,\nn\\
&(\psi^\dagger_2\psi_1\psi^\dagger_1\psi_1)(\vec R)=-\frac{n^2}{4}\Big(1+\sqrt{1-\frac{\kappa_1^2}{\kappa_0^2}}\cos(2\kappa_1x-\delta)\Big)\times\nn\\
&\times\Big(\cos(2\kappa_1x-\delta)+\sqrt{1-\frac{\kappa_1^2}{\kappa_0^2}}+\frac{i\kappa_1}{\kappa_0}\sin(2\kappa_1x-\delta)\Big)\,,\nn\\
&(\psi^\dagger_2\psi_1\psi^\dagger_2\psi_2)(\vec R)=-\frac{n^2}{4}\Big(1+\sqrt{1-\frac{\kappa_1^2}{\kappa_0^2}}\cos(2\kappa_1x-\delta)\Big)\times\nn\\
&\times\Big(\cos(2\kappa_1x-\delta)+\sqrt{1-\frac{\kappa_1^2}{\kappa_0^2}}+\frac{i\kappa_1}{\kappa_0}\sin(2\kappa_1x-\delta)\Big)\,.\nn\\
\end{align}
In the above, $\kappa_1=\kappa_0\sqrt{1-\frac{\Omega^2}{4(G_1+\kappa_0^2)^2}}$. It is interesting to note that at the phase boundary $\Omega=2\sqrt{(\kappa_0^2+G_1)(\kappa_0^2-2G_2)\frac{2G_2}{G_1+2G_2}}$, where the stripe phase goes to the plane wave phase, the contact densities are discontinuous, i.e., they do not continuously match to their values in the two phases. In contrast, the contact density in the plane wave phase continuously approaches its value in the zero momentum phase at the phase boundary $\Omega=2(\kappa_0^2-2G_2)$. Thus, we also observe the effect of the nature of the phase transition on the contact densities.
\section{Conclusions}\label{conclusionssec}
Ultra-cold bosonic systems provide a rich platform to explore various aspects of Bose-Einstein condensation. The spin-orbit coupled bosonic system is one such platform. Such a system is very versatile, and one can study it both theoretically and experimentally by tuning the various externally applied fields. In this paper, we have considered one such case where the pseudo-spinor atoms are subject to externally applied Raman lasers counterpropagating along the x-axis. 

In the present study, we have investigated the asymptotic fall-offs of the momentum-space distribution function in the spin-orbit and coherently coupled bosonic systems. It has been shown that such a fall-off determines universal behaviour of the ultra-cold atomic systems.
These studies have been carried out extensively in the fermionic case; however, we are not aware of any such study, at least explicitly, that has been carried out in the spin-orbit and coherently coupled bosonic systems. The present paper aims to fill this gap.

The leading fall-off behaviour can be extracted from the knowledge of the operator product expansion (OPE) of a pair of atomic fields. We have derived the OPE and expressed the product of two operators $\psi_{\sigma'}$ and $\psi^\dagger_\sigma$ as a series expansion in terms of local operators of the theory. We have restricted ourselves to the contact density term whose coefficient is non-analytic in $\vec r$. As Tan and Braaten et al. have shown, the contact density controls the large-$k$ universal behaviour in a quantum-mechanical system. It is important to emphasize that we have not considered here the three-body physics which is called the Efimov effect. The Efimov effect becomes important in the bosonic system, and as shown in~\cite{Smith:2013eoa}, taking it into account introduces new contact densities. It will be important to investigate this in the spin-orbit coupled system.

One important question which we want to leave for future work is the OPE of the conserved currents. It is important to explore whether there are universal physics in the large $ k$-behaviour of the OPE of the currents. For this, one can start with the fermionic systems and then move to the bosonic systems.

\begin{acknowledgments}
R Gupta would like to thank Sandeep Gautam for  useful discussions and his comments on the draft. The work of R Gupta is supported by SERB grant CRG/2023/001388.
\end{acknowledgments}
\begin{widetext}

\appendix

\section{One loop Computation}\label{OneLoopCompt.}
Fourier transform
\bea
\psi_\sigma(t,\vec r)=\int\frac{dEd^3k}{(2\pi)^4}e^{-iEt-i\vec k\cdot\vec r}\psi_\sigma(E,\vec k)
\eea
In this appendix, we state here the result of the one loop integration. We start with $d=3$ spatial dimensions. We have
\bea
\mathcal Z_1(E)&=&\int\frac{de\,d^3q}{(2\pi)^4}G_{11}(e,\vec q)G_{11}(E-e,-\vec q)\nn\\
&=&\int\frac{d^3q}{(2\pi)^3}\frac{i}{E-q^2-\kappa_0^2+i\epsilon}+\frac{i\Omega^2}{2}\int\frac{d^3q}{(2\pi)^3}\frac{1}{(E-q^2-\kappa_0^2+i\epsilon)((E-q^2-\kappa_0^2+i\epsilon)^2-\tilde\Omega^2)}\nn\\
&=&\frac{i}{4\pi}\sqrt{-E+\kappa_0^2-i\epsilon}+\frac{i\Omega^2}{2}I_1(E)
\eea
In the above $\tilde\Omega=\sqrt{\Omega^2+4\kappa_0^2q_x^2}$.
In the last step, we have used the dimensional regularization to evaluate the first integral. In the second integral, we have integrated over the transverse directions. We are finally leftover with the integration along $x$-component of the momentum.

Similarly,
\begin{align}
&\mathcal Z_2(E)=\int\frac{de\,d^3q}{(2\pi)^4}G_{11}(e,\vec q)G_{12}(E-e,-\vec q)
=\frac{i\Omega}{2}\int\frac{d^3q}{(2\pi)^3}\frac{1}{(E-q^2-\kappa_0^2+i\epsilon)^2-\tilde\Omega^2}=\frac{i\Omega}{2}I_2(E)\\
&\mathcal Z_3(E)=\int\frac{de\,d^3q}{(2\pi)^4}G_{12}(e,\vec q)G_{12}(E-e,-\vec q)=\frac{i\Omega^2}{2}I_1(E)\nn\\
&\mathcal Z_4(E)=\int\frac{de\,d^3q}{(2\pi)^4}G_{11}(e,\vec q)G_{22}(E-e,-\vec q)=i\int\frac{d^3q}{(2\pi)^3}\frac{1}{E-q^2-\kappa_0^2+i\epsilon}\Big[1+\frac{\frac{\Omega^2}{2}+4\kappa_0^2q_x^2}{(E-q^2-\kappa_0^2+i\epsilon)^2-\tilde\Omega^2}\Big]\\
&\qquad\quad=\frac{i}{4\pi}\sqrt{-E+\kappa_0^2-i\epsilon}+i\int\frac{d^3q}{(2\pi)^3}\frac{1}{E-q^2-\kappa_0^2+i\epsilon}\Big[\frac{\frac{\Omega^2}{2}+4\kappa_0^2q_x^2}{(E-q^2-\kappa_0^2+i\epsilon)^2-\tilde\Omega^2}\Big]\nn\\
&\qquad\quad=\frac{i}{4\pi}\sqrt{-E+\kappa_0^2-i\epsilon}+i\frac{\Omega^2}{2}I_1(E)+4i\kappa_0^2I_3(E)\,.
\end{align}
Let us evaluate each of the above integrals $I_i(E)$. We will work in the complex $E$-plane in the branch defined by $\text{Arg}(E-\kappa_0^2+i\epsilon)\in [-\pi,\pi)$. Then, we get
\begin{align}
&I_1(E)=-\frac{1}{32\pi^{\frac{3}{2}}\kappa_0^2\sqrt{-E+\kappa_0^2-i\epsilon}}\sum_{n=1}^\infty\frac{\Gamma(n-\frac{1}{2})}{\Gamma(n+1)}\Big(\frac{\kappa_0^2}{-E+\kappa_0^2-i\epsilon}\Big)^n{}_2F_1(1-n,n+\frac{1}{2},\frac{3}{2}-n,\frac{\Omega^2}{4\kappa_0^2(-E+\kappa_0^2-i\epsilon)})\,,\nn\\
&I_2(E)=\frac{1}{8\pi^{\frac{3}{2}}\kappa_0^2\sqrt{-E+\kappa_0^2-i\epsilon}}\sum_{n=0}^\infty\frac{\Gamma(n+\frac{1}{2})}{(2n+1)\Gamma(n+1)}\Big(\frac{\kappa_0^2}{-E+\kappa_0^2-i\epsilon}\Big)^n{}_2F_1(-n,n+\frac{1}{2},\frac{1}{2}-n,\frac{\Omega^2}{4\kappa_0^2(-E+\kappa_0^2-i\epsilon)})\,,\nn\\
&I_3(E)=-\frac{1}{32\pi^{\frac{3}{2}}\sqrt{-E+\kappa_0^2-i\epsilon}}\sum_{n=0}^\infty\frac{\Gamma(n+\frac{1}{2})}{\Gamma(n+2)}\Big(\frac{\kappa_0^2}{-E+\kappa_0^2-i\epsilon}\Big)^n{}_2F_1(-n,n+\frac{1}{2},-\frac{1}{2}-n,\frac{\Omega^2}{4\kappa_0^2(-E+\kappa_0^2-i\epsilon)})\,.\nn\\
\end{align}
These integrals simplify considerably in the coherently coupled case. Setting $\kappa_0=0$, we obtain
\begin{align}
&Z_1(E)=\frac{i}{4\pi}\sqrt{-E-i\epsilon}+\frac{1}{16\pi}\Big[-2\sqrt{E+i\epsilon}+\sqrt{E+i\epsilon+\Omega}+\sqrt{E+i\epsilon-\Omega}\Big]\,.\nn\\
&Z_2(E)=-\frac{1}{16\pi}(\sqrt{E+i\epsilon+\Omega}-\sqrt{E+i\epsilon-\Omega})\,.\nn\\
&Z_3(E)=\frac{1}{16\pi}\Big[-2\sqrt{E+i\epsilon}+\sqrt{E+i\epsilon+\Omega}+\sqrt{E+i\epsilon-\Omega}\Big]\,.\nn\\
&Z_4(E)=\frac{i}{4\pi}\sqrt{-E-i\epsilon}+\frac{1}{16\pi}\Big[-2\sqrt{E+i\epsilon}+\sqrt{E+i\epsilon+\Omega}+\sqrt{E+i\epsilon-\Omega}\Big]\,.
\end{align}
\section{Expectation value of the bilocal operator}\label{ProdExpGreensFn}
In this section, we present the integral expressions that appears in the computation of the expectation value of the bilocal operator. These are
\begin{align}
&1.\,\,\int\frac{de\,d^3p}{(2\pi)^4}e^{i\vec p\cdot\vec r}G_{11}^2(\vec p,e)G_{11}(-\vec p,E-e)=\int\frac{d^3p}{(2\pi)^3}\frac{e^{i\vec p\cdot\vec r}}{(E-p^2-\kappa_0^2+i\epsilon)^2}\Big[-1-\frac{\Omega^2}{(E-p^2-\kappa_0^2+i\epsilon)^2-\tilde\Omega^2}\nn\\
&\qquad\qquad\qquad\qquad\qquad\qquad\qquad\qquad+\frac{\Omega^2}{4\tilde\Omega}\Big\{\frac{(-\frac{\tilde\Omega}{2}+p_x\kappa_0)}{(E-p^2-\kappa_0^2+\tilde\Omega+i\epsilon)^2}-\frac{(\frac{\tilde\Omega}{2}+p_x\kappa_0)}{(E-p^2-\kappa_0^2-\tilde\Omega+i\epsilon)^2}\Big\}\Big]\label{FirstIntegral}\\\nn\\
&2.\,\,\int\frac{de\,d^3p}{(2\pi)^4}e^{i\vec p\cdot\vec r}G_{12}^2(\vec p,e)G_{22}(-\vec p,E-e)=\frac{\Omega^2}{4}\int\frac{d^3q}{(2\pi)^3}\frac{e^{i\vec q\cdot\vec r}}{(E-q^2-\kappa_0^2+i\epsilon)^2}\Big[\frac{p_x\kappa_0-\frac{\tilde\Omega}{2}}{\tilde\Omega(E-q^2-\kappa_0^2+i\epsilon+\tilde\Omega)^2}\nn\\
&\qquad\qquad\qquad\qquad\qquad\qquad\qquad\qquad\qquad\qquad\qquad\qquad\qquad-\frac{p_x\kappa_0+\frac{\tilde\Omega}{2}}{\tilde\Omega(E-q^2-\kappa_0^2+i\epsilon-\tilde\Omega)^2}\Big]\\\nn\\
&3.\,\,\int\frac{ded^3q}{(2\pi)^4}e^{i\vec q\cdot\vec r}G_{11}(e,\vec q)G_{12}(e,\vec q)G_{12}(E-e,-\vec q)=\frac{\Omega^2}{4}\int\frac{d^3q}{(2\pi)^3}\frac{e^{i\vec q\cdot\vec r}}{\tilde\Omega(E-q^2-\kappa_0^2+i\epsilon)^2}\Big[-\frac{2\tilde\Omega}{(E-q^2-\kappa_0^2+i\epsilon)^2-\tilde\Omega^2}\nn\\
&\qquad\qquad\qquad\qquad\qquad\qquad\qquad\qquad\qquad+\frac{\kappa_0q_x-\frac{\tilde\Omega}{2}}{(E-q^2-\kappa_0^2+\tilde\Omega+i\epsilon)^2}-\frac{\kappa_0q_x+\frac{\tilde\Omega}{2}}{(E-q^2-\kappa_0^2-\tilde\Omega+i\epsilon)^2}\Big]\\\nn\\
&4.\,\,\int\frac{ded^3q}{(2\pi)^4}e^{i\vec q\cdot\vec r}G^2_{11}(e,\vec q)G_{22}(E-e,-\vec q)=\int\frac{d^3q}{(2\pi)^3}\frac{e^{i\vec q\cdot\vec r}}{(E-q^2-\kappa_0^2+i\epsilon)^2}\Big[-1+\frac{(\kappa_0q_x-\frac{\tilde\Omega}{2})^3}{\tilde\Omega(E-q^2-\kappa_0^2+\tilde\Omega+i\epsilon)^2}\nn\\
&\qquad\qquad\qquad\qquad+\frac{2(\kappa_0q_x-\frac{\tilde\Omega}{2})^2}{\tilde\Omega(E-q^2-\kappa_0^2+\tilde\Omega+i\epsilon)}-\frac{(\kappa_0q_x+\frac{\tilde\Omega}{2})^3}{\tilde\Omega(E-q^2-\kappa_0^2-\tilde\Omega+i\epsilon)^2}-\frac{2(\kappa_0q_x+\frac{\tilde\Omega}{2})^2}{\tilde\Omega(E-q^2-\kappa_0^2-\tilde\Omega+i\epsilon)}\Big]\\\nn\\
&5.\,\,\int\frac{ded^3q}{(2\pi)^4}e^{i\vec q\cdot\vec r}G^2_{12}(e,\vec q)G_{12}(E-e,-\vec q)=-\frac{\Omega^3}{2}\int\frac{d^3p}{(2\pi)^3}e^{i\vec q\cdot\vec r}\frac{1}{E-p^2-\kappa_0^2+i\epsilon}\frac{1}{[(E-p^2-\kappa_0^2+i\epsilon)^2-\tilde\Omega^2]^2}\\\nn\\
&6.\,\,\int\frac{ded^3q}{(2\pi)^4}e^{i\vec q\cdot\vec r}G_{11}(e,\vec q)G_{12}(e,\vec q)G_{22}(E-e,-\vec q)=-\frac{\Omega}{2}\int\frac{d^3q}{(2\pi)^3}e^{i\vec q\cdot\vec r}\frac{(E-q^2-\kappa_0^2+2q_x\kappa_0)^2}{(E-q^2-\kappa_0^2+i\epsilon)[(E-q^2-\kappa_0^2+i\epsilon)^2-\tilde\Omega^2]^2}\\\nn\\
&7.\,\,\int\frac{ded^3q}{(2\pi)^4}e^{i\vec q\cdot\vec r}G_{11}(e,\vec q)G_{12}(e,\vec q)G_{11}(E-e,-\vec q)=-\frac{\Omega}{2}\int\frac{d^3p}{(2\pi)^3}\frac{e^{i\vec p\cdot\vec r}}{(E-p^2-\kappa_0^2+i\epsilon)^2}\Big[\frac{\Omega^2(E-p^2-\kappa_0^2)}{\Big[(E-p^2-\kappa_0^2+i\epsilon)^2-\tilde\Omega^2\Big]^2}\nn\\
&\qquad\qquad\qquad\qquad\qquad\qquad\qquad\qquad\qquad\qquad\qquad\qquad+\frac{(E-p^2-\kappa_0^2-2q_x\kappa_0)}{(E-p^2-\kappa_0^2+i\epsilon)^2-\tilde\Omega^2}\Big]\\\nn\\
&8.\,\,\int\frac{ded^3q}{(2\pi)^4}e^{i\vec q\cdot\vec r}G^2_{11}(e,\vec q)G_{12}(E-e,-\vec q)=-\frac{\Omega}{2}\int\frac{d^3p}{(2\pi)^3}e^{i\vec p\cdot\vec r}\frac{(E-q^2-\kappa_0^2+2q_x\kappa_0)}{(E-q^2-\kappa_0^2+i\epsilon)^2((E-q^2-\kappa_0^2+i\epsilon)^2-\tilde\Omega^2)}\times\nn\\
&\qquad\qquad\qquad\qquad\qquad\qquad\qquad\qquad\qquad\qquad\times\Big[2+\frac{2(E-q^2-\kappa_0^2)\kappa_0q_x+\tilde\Omega^2}{(E-q^2-\kappa_0^2+i\epsilon)^2-\tilde\Omega^2}\Big]
\end{align}
Notice the organization of the above integrals. The first term in the first line of the integral~\eqref{FirstIntegral} scales as $\frac{1}{\Lambda}$ when we scale $\vec p\sim\Lambda \vec p$. On the other hand, the rest of the integrals scale as $\frac{1}{\Lambda^5}$. One can look at the similar scalings of the other integrals. These scalings are important to determine the presence of the non-analytic term in $r$.

\section{Coherently coupled case}\label{ProdGreensFnCohenertlyCoupled}
The above integrals, involved in computing the OPE, are easily computable for the coherent case, i.e. $\kappa_0=0$. We compute these integrals in the perturbation expansion in $r$ and keep terms leading order in $r$. The integrals are
\bea
\int\frac{ded^3p}{(2\pi)^4}e^{i\vec p\cdot\vec r}\,G_{11}(e,\vec p)^2G_{11}(E-e,-\vec p)=-\frac{1}{8\pi\Omega}\Big(\sqrt{-E-i\epsilon+\Omega}-\sqrt{-E-i\epsilon-\Omega}\Big)+\frac{r}{8\pi}+\mathcal O(r^2)
\eea
\bea
\int\frac{ded^3q}{(2\pi)^4}e^{i\vec q\cdot\vec r}G_{11}(e,\vec q)G_{12}(e,\vec q)G_{11}(E-e,-\vec q)=-\frac{1}{64\pi}\Big(\frac{1}{\sqrt{-E-i\epsilon+\Omega}}-\frac{1}{\sqrt{-E-i\epsilon-\Omega}}\Big)+\mathcal O(r^2)
\eea
\bea
\int\frac{ded^3q}{(2\pi)^4}e^{i\vec q\cdot\vec r}G^2_{11}(e,\vec q)G_{12}(E-e,-\vec q)=\frac{1}{64\pi\Omega}\Big(8\sqrt{-E-i\epsilon}+\frac{4E+5\Omega}{\sqrt{-E-i\epsilon-\Omega}}+\frac{4E-5\Omega}{\sqrt{-E-i\epsilon+\Omega}}\Big)+\mathcal O(r^2)
\eea
\beq
\int\frac{ded^3q}{(2\pi)^4}e^{i\vec q\cdot\vec r}G_{11}(e,\vec q)G_{12}(e,\vec q)G_{12}(E-e,-\vec q)=\frac{1}{16\pi\Omega}\Big(\frac{\Omega}{\sqrt{-E-i\epsilon}}+\sqrt{-E-i\epsilon-\Omega}-\sqrt{-E-i\epsilon+\Omega}\Big)+\mathcal O(r^2)
\eeq
\beq
\int\frac{de\,d^3p}{(2\pi)^4}e^{i\vec p\cdot\vec r}G_{12}^2(\vec p,e)G_{22}(-\vec p,E-e)=\frac{1}{64\pi\Omega}\Big[-\frac{2\Omega}{\sqrt{-E-i\epsilon}}+\frac{-4E+3\Omega}{\sqrt{-E-i\epsilon+\Omega}}+\frac{4E+3\Omega}{\sqrt{-E-i\epsilon-\Omega}}\Big]+\mathcal O(r^2)
\eeq

\beq
\int\frac{ded^3q}{(2\pi)^4}e^{i\vec q\cdot\vec r}G^2_{12}(e,\vec q)G_{12}(E-e,-\vec q)=-\frac{1}{64\pi\Omega}\Big[8\sqrt{-E-i\epsilon}+\frac{4E-3\Omega}{\sqrt{-E-i\epsilon+\Omega}}+\frac{4E+3\Omega}{\sqrt{-E-i\epsilon-\Omega}}\Big]+\mathcal O(r^2)
\eeq
\section{Expectation value of the Bi-local operator}\label{AppendixExpectationaValueOPE}
In this appendix, we present the results for the expectation value of the linear combination of the contact operators $\mu_1(\psi^\dagger_1\psi_1)^2+\mu_2(\psi^\dagger_2\psi_2)^2+\mu_3\psi^\dagger_1\psi_1\psi^\dagger_2\psi_2$ between two body state. For the states $i,j\rightarrow k,\ell$, there are three diagrams that contribute to the expectation value: one involving no scattering, 2nd involves single scattering and third has two scattering. The contributions are 
\begin{align}\label{ExpectationaValueOPE.1}
&4\m_1\delta_{i1}\delta_{j1}\delta_{k1}\delta_{\ell 1}+4\m_2\delta_{i2}\delta_{j2}\delta_{k2}\delta_{\ell 2}+\m_3(\delta_{i1}\delta_{j2}+\delta_{i2}\delta_{j1})(\delta_{k1}\delta_{\ell 2}+\delta_{k2}\delta_{\ell 1})\nn\\
&+2\int\frac{ded^3p}{(2\pi)^4}\Big[\m_1\delta_{i1}\delta_{j1}G_{1u}(e,\vec p)G_{1v}(E-e,-\vec p)(i\mathcal M_{uv\rightarrow k\ell})+\m_2\delta_{i2}\delta_{j2}G_{2u}(e,\vec p)G_{2v}(E-e,-\vec p)(i\mathcal M_{uv\rightarrow k\ell})\nn\\
&+\frac{\m_3}{2}(\delta_{i1}\delta_{j2}+\delta_{i2}\delta_{j1})G_{1u}(e,\vec p)G_{2v}(E-e,-\vec p)(i\mathcal M_{uv\rightarrow k\ell})\Big]+2\int\frac{ded^3p}{(2\pi)^4}\Big[\m_1(i\mathcal M_{ij\rightarrow uv})\delta_{1k}\delta_{1\ell}G_{u1}(e,\vec p)G_{v1}(E-e,-\vec p)\nn\\
&+\m_2(i\mathcal M_{ij\rightarrow uv})\delta_{k2}\delta_{\ell2}G_{u2}(e,\vec p)G_{v2}(E-e,-\vec p)+\frac{\m_3}{2}(i\mathcal M_{ij\rightarrow uv})G_{u1}(e,\vec p)G_{v2}(E-e,-\vec p)(\delta_{1k}\delta_{2\ell}+\delta_{1\ell}\delta_{2k})\Big]\nn\\
&+\int\prod_{i=1,2}\frac{de_id^3p_i}{(2\pi)^4}\Big[\m_1(i\mathcal M_{ij\rightarrow uv})G_{u1}(e_1,\vec p_1)G_{v1}(E-e_1,-\vec p_1)G_{1p}(e_2,\vec p_2)G_{1q}(E-e_2,-\vec p_2)(i\mathcal M_{pq\rightarrow k\ell})\nn\\
&+\m_2(i\mathcal M_{ij\rightarrow uv})G_{u2}(e_1,\vec p_1)G_{v2}(E-e_1,-\vec p_1)G_{2p}(e_2,\vec p_2)G_{2q}(E-e_2,-\vec p_2)(i\mathcal M_{pq\rightarrow k\ell})\nn\\
&+\m_3(i\mathcal M_{ij\rightarrow uv})G_{u1}(e_1,\vec p_1)G_{v2}(E-e_1,-\vec p_1)G_{1p}(e_2,\vec p_2)G_{2q}(E-e_2,-\vec p_2)(i\mathcal M_{pq\rightarrow k\ell})\Big]
\end{align}
Similarly, repeating the above calculation for the series of the operators $\eta_1\psi^\dagger_2\psi_1\psi^\dagger_1\psi_1+\eta_2\psi^\dagger_2\psi_1\psi^\dagger_2\psi_2+\eta_3\psi^\dagger_1\psi_2\psi^\dagger_1\psi_1+\eta_4\psi^\dagger_1\psi_2\psi^\dagger_2\psi_2$, we obtain
\begin{align}\label{ExpectationaValueOPE.2}
&2\eta_1\delta_{i1}\delta_{j1}(\delta_{2k}\delta_{1\ell}+\delta_{1k}\delta_{2\ell})+\eta_2\delta_{2k}\delta_{\ell2}(\delta_{2i}\delta_{1j}+\delta_{1i}\delta_{2j})+2\eta_3\delta_{1k}\delta_{1\ell}(\delta_{1i}\delta_{2j}+\delta_{2i}\delta_{1j})+2\eta_4\delta_{i2}\delta_{j2}(\delta_{1k}\delta_{2\ell}+\delta_{2k}\delta_{1\ell})\nn\\
&+\int\frac{ded^3p}{(2\pi)^4}\,\Big[2\eta_1\delta_{i1}\delta_{j1}G_{2u}(e,\vec p)G_{1v}(E-e,-\vec p)+\eta_2(\delta_{i1}\delta_{j2}+\delta_{i2}\delta_{j1})G_{2u}(e,\vec p)G_{2v}(E-e,-\vec p)\nn\\
&+\eta_3(\delta_{i1}\delta_{j2}+\delta_{i2}\delta_{j1})G_{1u}(e,\vec p)G_{1v}(E-e,-\vec p)+2\eta_4\delta_{i2}\delta_{j2}G_{1u}(e,\vec p)G_{2v}(E-e,-\vec p)\Big](iM_{uv\rightarrow k\ell})\nn\\
&+\int\frac{ded^3p}{(2\pi)^4}\,(iM_{ij\rightarrow uv})\Big[\eta_1G_{u1}(e,\vec p)G_{v1}(E-e,-\vec p)(\delta_{1k}\delta_{2\ell}+\delta_{2k}\delta_{1\ell})+2\eta_2G_{u2}(e,\vec p)G_{v1}(E-e,-\vec p)\delta_{2k}\delta_{2\ell}\nn\\
&+2\eta_3G_{u2}(e,\vec p)G_{v1}(E-e,-\vec p)\delta_{1k}\delta_{1\ell}+\eta_4G_{u2}(e,\vec p)G_{v2}(E-e,-\vec p)(\delta_{1k}\delta_{2\ell}+\delta_{2k}\delta_{1\ell})\Big]\nn\\
&+\int\prod_{i=1,2}\frac{de_id^3p_i}{(2\pi)^4}\,(iM_{ij\rightarrow uv})\Big[\eta_1G_{u1}(e_1,\vec p_1)G_{v1}(E-e_1,-\vec p_1)G_{2p}(e_2,\vec p_2)G_{1q}(E-e_2,-\vec p_2)\nn\\
&+\eta_2G_{u1}(e_1,\vec p_1)G_{v2}(E-e_1,-\vec p_1)G_{2p}(e_2,\vec p_2)G_{2q}(E-e_2,-\vec p_2)+\eta_3G_{u2}(e_1,\vec p_1)G_{v1}(E-e_1,-\vec p_1)G_{1p}(e_2,\vec p_2)G_{1q}(E-e_2,-\vec p_2)\nn\\
&+\eta_4G_{u2}(e_1,\vec p_1)G_{v2}(E-e_1,-\vec p_1)G_{1p}(e_2,\vec p_2)G_{2q}(E-e_2,-\vec p_2)\Big](iM_{pq\rightarrow k\ell})
\end{align}

\end{widetext}

\end{document}